\documentstyle[epsfig]{jaa}

\DeclareMathAlphabet{\mathsc}{OT1}{cmr}{m}{sc}
\def\testbx{bx}%
\DeclareRobustCommand{\ion}[2]{%
\relax\ifmmode
\ifx\testbx\f@series
{\mathbf{#1\,\mathsc{#2}}}\else
{\mathrm{#1\,\mathsc{#2}}}\fi
\else\textup{#1\,{\mdseries\textsc{#2}}}%
\fi}

\def\eg{{\it e.g.,~}}
\def\ie{{\it i.e.,~}}
\def\ltsim{\; \raise0.3ex\hbox{$<$\kern-0.75em \raise-1.1ex\hbox{$\sim$}}\; }
\def\gtsim{\; \raise0.3ex\hbox{$>$\kern-0.75em \raise-1.1ex\hbox{$\sim$}}\; }
\begin{document}
\title[Radio Halos and cluster mergers]
{The connection between radio halos and cluster mergers and the statistical
properties of the radio halo population}
\author[Cassano, Brunetti, Venturi] 
{R. Cassano $^{1,2}$ \thanks{e-mail: rcassano@ira.inaf.it}, G. Brunetti $^2$, T. Venturi$^2$\\
$^1$ Dipartimento di Astronomia, Universit\`a di Bologna, via Ranzani 1,  40127, Bologna, Italy \\
$^2$ IRA-INAF, via Gobetti 101, 40129 Bologna, Italy.}


\pubyear{xxxx}
\volume{xx}
\date{Received xxx; accepted xxx}
\maketitle
\label{firstpage}
\begin{abstract}

We discuss the statistical properties of the radio halo population in galaxy clusters. Radio bi-modality is observed in galaxy clusters: a fraction of clusters host giant radio halos while the majority of clusters do not show evidence of diffuse cluster-scale radio emission. The radio bi-modality has a correspondence in terms of dynamical state of the hosting clusters showing that merging clusters host radio halos and follow the well known radio--X-ray correlation, while more relaxed clusters do not host radio halos and populate a region well separated from that correlation. These evidences can be understood in the framework of a scenario where merger-driven turbulence re-accelerate the radio emitting electrons. We discuss the main statistical expectations of this scenario underlining the important role of upcoming LOFAR surveys to test present models.

\end{abstract} 

\begin{keywords}
galaxy: clusters: general -- Radio continuum: general -- X-rays: general -- Radiation mechanisms: non-thermal

\end{keywords}

\section{Introduction}

Radio and X-ray observations of galaxy clusters prove that thermal 
and non-thermal components coexist in the intra-cluster medium (ICM). 
While X-ray observations reveal thermal emission from diffuse hot gas, 
radio observations of an increasing number of galaxy clusters 
unveil the presence of ultra-relativistic particles and magnetic fields 
through the detection of diffuse, giant Mpc-scale synchrotron 
{\it radio halos} (RH) and {\it radio relics} (\eg Ferrari et al. 2008, Cassano 2009). 
RH are the most spectacular evidence of non-thermal components in the ICM. 
They are giant radio sources located in the cluster central regions, 
with spatial extent similar to that of the hot ICM and steep radio 
spectra, $\alpha> 1.1$  (\eg Venturi 2011).

There are well known correlations between the synchrotron monochromatic 
radio luminosity of RH ($P_{1.4}$) and the host cluster X-ray luminosity ($L_X$), mass and 
temperature (\eg Liang 2000; Feretti 2003; Cassano et al. 2006; Brunetti et al. 2009).
The most powerful RH are found in the most X-ray luminous, massive and 
hot clusters. These correlations suggest a close link between the 
non-thermal and the thermal/gravitational cluster physics.  

Most important, RH are presently found only in clusters that show recent/ongoing merging activity. 
To this regard, Buote (2001) provided the first quantitative 
comparison of the dynamical states of clusters with RH and the properties
of RH. He discovered a correlation between $P_{1.4}$  and the magnitude of the dipole power ratio 
$P_1/P_0$:  the more powerful RH are hosted in clusters
that experience the largest departures from virialization.

The RH-merger connection and the thermal--non-thermal correlations suggest
that the gravitational process of cluster formation may provide the energy 
to generate the non-thermal components in clusters through the acceleration 
of high-energy particles via shocks and turbulence (\eg Sarazin 2004,
Brunetti 2011). 

The origin of RH is still debated. One possibility is that RH are due
to synchrotron emission from secondary electrons generated by p-p collisions (\eg Dennison 1980),
in which case clusters must be 
gamma ray emitters due to the decay of the $\pi^0$ produced by the same collisions. However, the non-detections
of nearby galaxy clusters at GeV energies by FERMI puts strong constraints to the contribution of secondary
electrons to the non-thermal emission (Ackermann et al. 2010). Most important, the spectral and
morphological properties of a number of well studied RH appear inconsistent with a pure
hadronic origin of the emitting particles (\eg Brunetti et al 2008, 2009; Donnert et al 2010;
Macario et al 2010; Brown \& Rudnick 2011).

A second hypothesis is based on turbulent re-acceleration of relativistic particles in connection with
cluster-mergers events (\eg Brunetti et al. 2001; Petrosian 2001). This model received recently support from the discovery 
of RH with very steep spectra (\eg Brunetti et al 2008; Macario et al 2010). Future low-frequency radio telescopes (such as LOFAR and LWA) 
have the potential to test this scenario and to further explore the connection between RH and the process 
of cluster formation. 

Here we focus on the most recent advances in the study of the statistical propertis of RH and their connection 
with cluster mergers, and discuss the importance of future surveys at low radio frequencies to test present models.

\section{The GMRT RH Survey}
\subsection{The radio bi-modality of clusters}

An important step forward in our understanding of the statistical properties of RH and of
their connection with the process of cluster formation has been recently carried out thanks
to the Giant Metrewave Radio Telescope ({\it GMRT}) RH Survey (Venturi et al. 2007, 2008),
a deep observational campaign of a complete sample of X-ray selected galaxy clusters (with
X-ray luminosity$\geq 5\times 10^{44}$ erg/s in the redshift range $0.2-0.4$) performed 
at 610 MHz with the GMRT.

\begin{figure}
\centering
\includegraphics[width=5.2cm, angle=0]{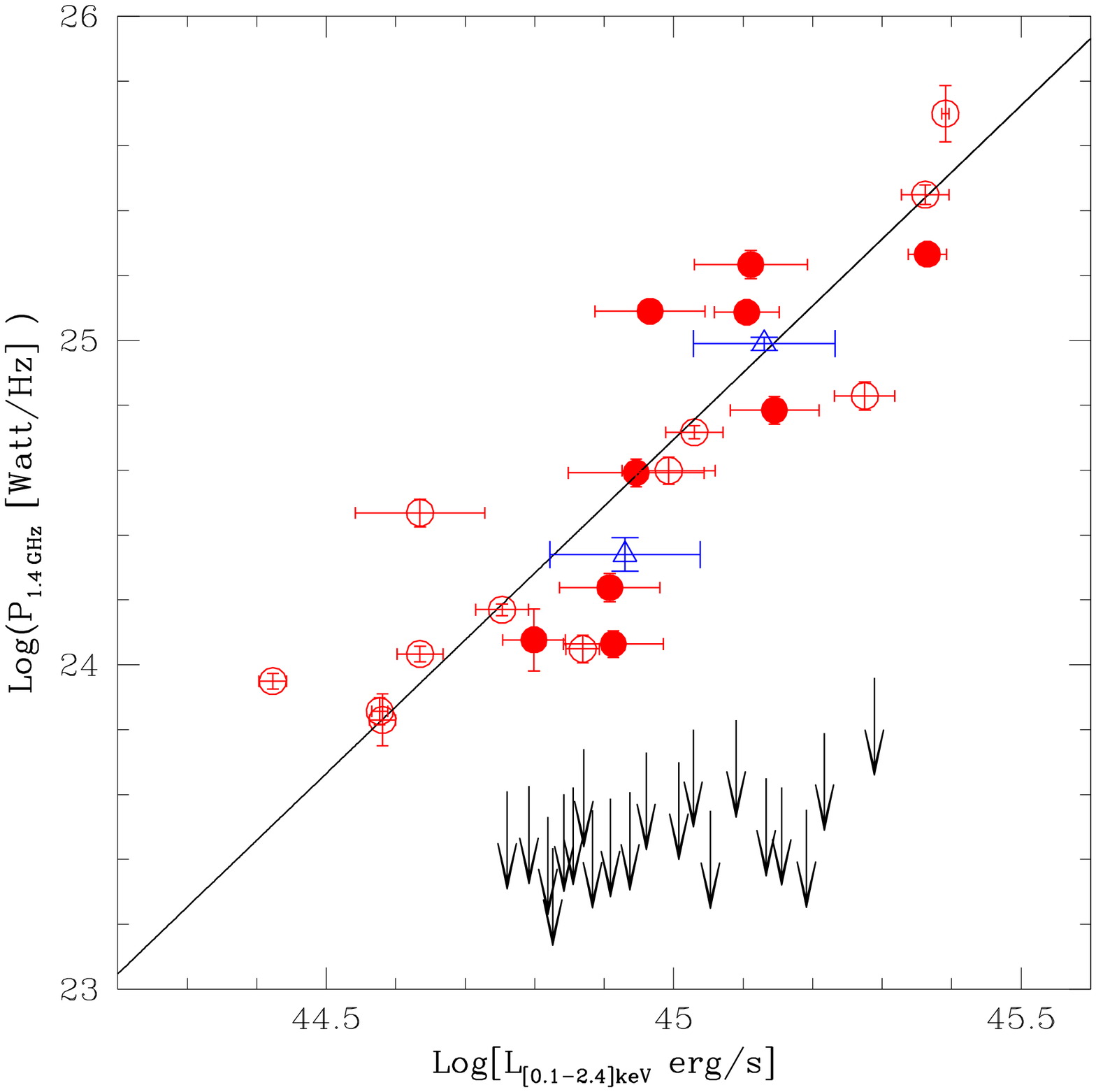}
\includegraphics[width=7.cm, angle=0]{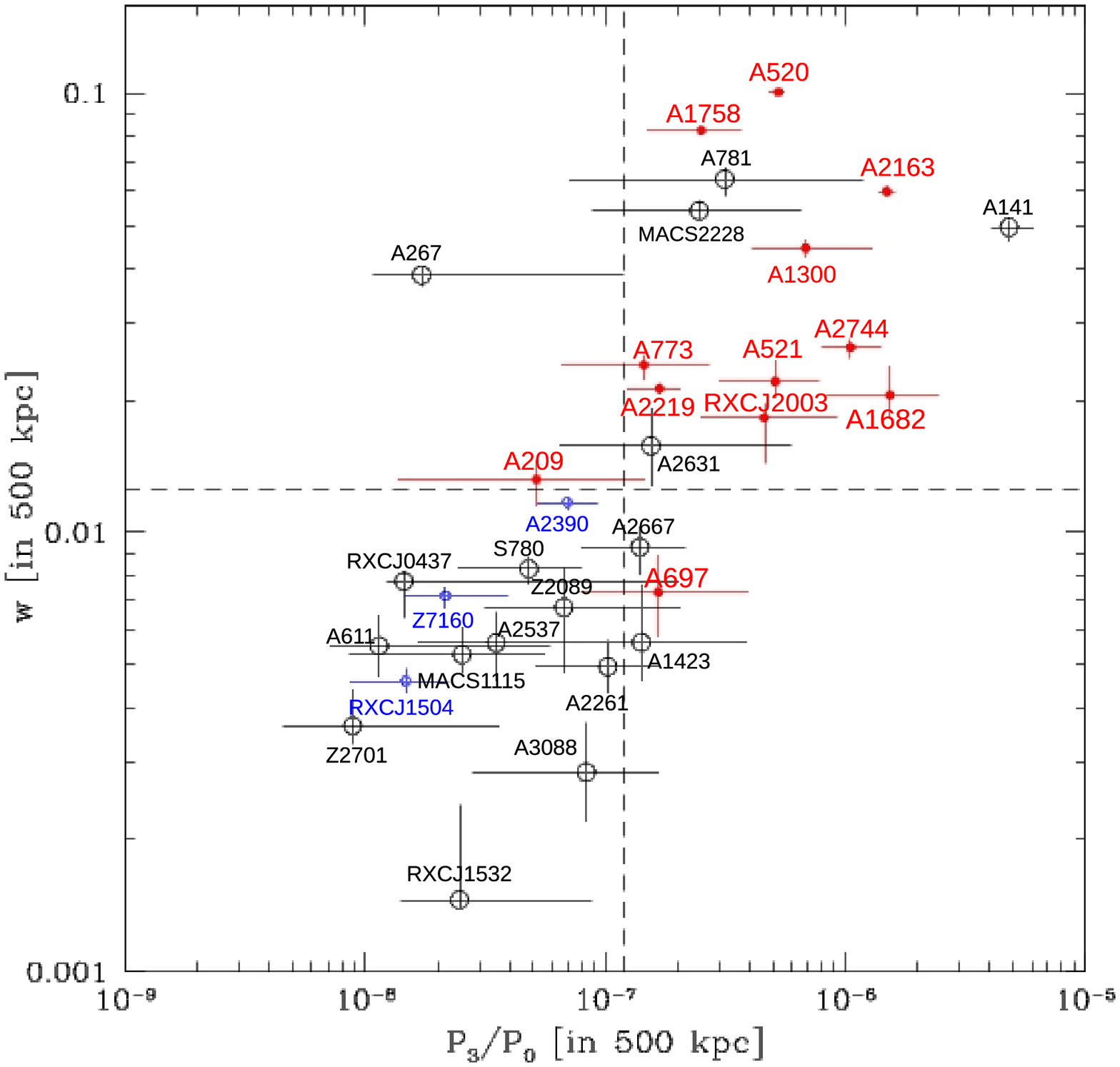}
\caption{\footnotesize
{\bf Left Panel}: distribution of clusters in the plane $P_{1.4}$ --  $L_{[0.1-2.4] keV}$ for clusters of 
the GMRT RH Survey (RH: red filled dots; non-RH: black arrows; mini-halos: blue symbols) 
and for RH from the literature (red open dots).
{\bf Right Panel}:  centroid shift {\it w} vs. {\it P3/P0}. 
Symbols are: RH (red filled dots), non-RH (black open dots), mini-halos (blue open dots). 
Vertical and horizontal dashed lines mark : $w = 0.012$ and $P_3/P_0=1.2\times10^{-7}$.}
\label{LrLx_sub}
\end{figure}

These observations allowed {\it for the first time} to prove statistically that diffuse cluster-scale 
radio emission is not ubiquitous in clusters: only ~30\% of the X-ray luminous 
($L_{X\,[0.1-2.4\,\mathrm{keV}]}\geq5\times10^{44}$ erg/s) clusters host a RH. 
Most important, it was possible to separate RH clusters from clusters without RH, showing a 
bimodal distribution of clusters in the $P_{1.4}-L_X$ diagram (Brunetti et al. 2007): 
RH trace the well known correlation between $P_{1.4}$ 
and $L_X$, while the upper limits to the radio luminosity of clusters without RH lie about one order 
of magnitude below that correlation (Fig.~\ref{LrLx_sub}, left panel). 
Why clusters with the same thermal X-ray luminosity (and at the same cosmological 
epoch) have different non-thermal properties ? 
Based on information from the literature available for a fraction of the clusters of the GMRT RH Survey, 
Venturi et al. (2008) suggested that the behavior of clusters in the 
$P_{1.4}-L_X$ diagram is connected with their dynamical state.

\subsection{The dynamical state of GMRT clusters}
\label{sec:dyn_cluster}

To test the connection between RH and cluster mergers Cassano et al. (2010a) using {\it Chandra} archive X-ray 
data of a sub-sample of GMRT clusters provided a quantitative measure of the degree of 
the cluster disturbance adopting three different methods: the power ratios 
(\eg Buote et al. 1995, Jeltema et al. 2005), the emission centroid shift (\eg Mohr et al. 1993, Poole et al. 2006), 
and the surface brightness concentration parameter (\eg Santos et al. 2008). 
A detailed descriptions of these measurements is given in Cassano et al. (2010a, and ref. therein).

They found a clear segregation between clusters with and without RH in terms of their dynamical state: RH are only found in dynamically
disturbed clusters (those with high values of $P_3/P_0$, $P_3/P_0 \gtsim 
1.2\times 10^{-7}$ and $w$, $w\gtsim0.012$, and low values of 
$c$, $c\ltsim0.2$), while clusters with no evidence of Mpc-scale synchrotron 
emission are more relaxed systems.  As an example in Fig.\ref{LrLx_sub} (right panel) 
we report the distribution of the clusters in the ($w,P_3/P_0$) plane.
This result was also tested quantitatively by running Monte Carlo simulations 
(see Cassano et al. 2010a for details) that proved that the observed 
distribution differs from a random one (\ie independent of cluster dynamics) 
at more than $4\sigma$. 

We note also that not all the disturbed systems host RH: specifically, we found 4 ``radio anomalies'' in Fig.\ref{LrLx_sub} (right panel), 
Abell 781, MACS 2228, Abell 141 and Abell 2631, \ie clusters that have the same morphological parameters ($P_3/P_0$, $w$ and $c$)
of clusters with RH but that do not host a RH.

\section{The evolution of RH in the $P_{1.4}-L_X$ diagram}

The radio bi-modality of galaxy clusters and the connection with their
dynamical state suggest the following coupled evolution between RH and clusters:

\begin{itemize}
\item[{\it a)}] clusters host RH for a period of time, in connection with
cluster mergers, and populate the $P_{1.4}-L_X$ correlation (Fig.~\ref{LrLx_sub}, left panel);

\item[{\it b)}] at later times, when clusters become dynamically relaxed, the Mpc-scale
synchrotron emission is gradually suppressed and clusters populate the region of the upper-limits.

\end{itemize}

19 clusters of the GMRT sample have $L_X\ge8.5\times10^{44}$ erg/s, in which case the radio power
of halos is $\sim$ 1 order of magnitude larger than the level of radio upper limits. Among these 19 clusters:
5 host giant RH, 11 are ``radio-quiet'' and only one is in the transition region (here not reported in 
Fig.~\ref{LrLx_sub}, left panel). This allows to estimate the lifetime of RH, $\tau_{RH}\approx 1$ Gyr, 
and the time clusters spend in the ``radio-quiet'' phase, $\tau_{rq}\approx 2-2.5$ Gyr (Brunetti et al 2009). 
Most important, at these luminosities, the ``empty'' region between RH and ``radio quiet'' clusters 
in the $P_{1.4}-L_X$ diagram constrains the time-scale of the evolution (suppression and amplification) of 
the synchrotron emission (Brunetti et al. 2007, 2009) to be much shorter than both the lifetime of clusters in the sample 
and the period of time clusters spend in the RH stage.  Monte Carlo analysis of the distribution of clusters in 
Fig.~\ref{LrLx_sub} (left panel) shows that the time interval that clusters spend in the ``empty'' region 
(thus the corresponding time-scale for amplification and suppression of RH) is $\tau_{evol} \approx 200$ Myr, 
with the probability that $\tau_{evol}$ is as large as 1 Gyr $\le$ 1\% (Brunetti et al. 2009).

The evolution of the radio properties of galaxy clusters in the plane $P_{1.4}-L_X$ is driven by the evolution 
of the relativistic components (B and particles) in the ICM. The tight constraints on the timeÐscale of this evolution, 
$\tau_{evol} \approx 200$ Myr, provides crucial information on the physics of the particle acceleration and magnetic field amplification.

\subsection{The role of cluster magnetic field}

A possible explanation of the bi-modality is that cluster mergers amplify the magnetic field
in the ICM leading to the amplification of the synchrotron emission on Mpc scales. In this case,
merging clusters hosting RH should have larger magnetic fields, $\delta B+B$, with the excess $\delta B$ 
being generated during mergers and then dissipated when clusters become ``radio quiet'' and 
dynamically more relaxed (Brunetti et al. 2007, 2009; Kushnir et al. 2009; Keshet \& Loeb 2010).

A magnetic field evolution was postulated to reconcile a secondary origin of RH with the observed bi-modality;
these models would indeed predict RH in all clusters, provided the ICM is magnetized at similar (few $\mu$G) level
\footnote{See, however, En\ss lin et al. (2011) where it is proposed a (ad hoc) super-Alfv\'enic diffusion of cosmic rays to have 
radio (and $\gamma$-rays) evolution driven by cosmic ray diffusion.}.

A suppression of a factor $\ge$ 10 in terms of synchrotron emission 
constrains the ratio $\delta B/B$. At $z\approx0.25$ (typical of GMRT clusters) in the case  $\delta B+B<<B_{cmb}$ (where $B_{cmb}= 
3.2(1+z)^2\,\mu$G is the equivalent magnetic field of the CMB) the energy density of the magnetic field in 
RH clusters should be $\ge10$ times larger than that in ``radio quiet'' clusters, and even larger ratios must 
be assumed if $\delta B+B>>B_{cmb}$ (Brunetti et al. 2009).

This significant difference between the magnetic field strength in RH and ``radio quiet'' clusters is a prediction of this scenario that,
however, is not supported by present observations.
Faraday Rotation measurements (RM) in galaxy clusters do not show any statistical difference between 
the energy density of the large scale (10-100 kpc coherent scales) magnetic field in RH clusters and that in ``radio quiet'' clusters (e.g., Carilli \& Taylor 2002). In a recent paper, Govoni et al (2010) studied the $\sigma_{RM}-S_{X}$ distribution
($\sigma_{RM}$ is the $\sigma$ of the RM and $S_X$ the thermal X-ray cluster brightness, see Govoni et al. 2010 for details) of radio sources in a sample of hot galaxy clusters, including both ``radio-quiet'' and RH clusters. They showed that all clusters follow the same  $\sigma_{RM}-S_{X}$ trend, and since $\sigma_{RM} \propto \Lambda_c\int (n_{th}B_{||})^2 dl$ ($\Lambda_c$ the field coherent scale) this allow to conclude that the magnetic field strength in ``radio-quiet'' and RH clusters is similar (see also Brunetti \& Cassano 2010). 

More recently, Bonafede et al. (2011) investigate the fractional polarization trends in 39 massive clusters with different non-thermal properties. 
They found no statistical evidence for a difference in the depolarization trends, concluding that there is no evidence for different magnetic fields in these clusters.
 
All these results suggest that the bi-modality in the $P_{1.4}-L_X$ plane cannot be attributed to a bi-modality in magnetic 
field properties.

\subsection{The role of relativistic particles}

Since present data suggest that the magnetic field is not the main responsible of the evolution of the radio properties of galaxy clusters, relativistic electrons must drive the generation and fading away of RH.

Turbulent acceleration models provide a natural way to explain the radio bi-modality of galaxy clusters and the fast evolution of RH. In these models relativistic electrons are re-accelerated {\it in situ} by turbulence on Mpc-scales during cluster mergers and cool as soon as the clusters become more relaxed due to the dissipation of turbulence. 
The cooling time of relativistic electrons emitting in the radio band is $\sim 10^8$ yrs (\eg Sarazin 1999) which is very short and consistent with (smaller than) the evolution timescale of RH constrained by the distribution of clusters in Fig.\ref{LrLx_sub} (left panel).

This scenario predicts that during a merger as soon as the turbulence reaches small (resonant) scales, particles are accelerated and generate synchrotron emission at GHz frequencies (the clusters move from the region of the upper limits to the $P_{1.4}-L_X$ correlation and should appear dynamically disturbed) within a timescale of  few $100$ Myrs. The process should persist for a few crossing times of the central cluster Mpc region, that is fairly consistent with the RH lifetime $\tau_{RH}\sim$ Gyr, this is the period the clusters would appear on the $P_{1.4}-L_X$ correlation. As soon as the turbulence starts to dissipate (at the end of the merging phase) the synchrotron power is suppressed and the synchrotron emission at higher frequencies will fall below the detection limit of radio observations (the clusters move again in the region of the upper limits and appear more relaxed at X-rays wavelength).
A significant suppression of the synchrotron luminosity occurs in a time-scale of the order of a turbulent-eddy turnover time, $\approx 10^8$ years, consistent with the fact that RH do not populate the region between the correlation and the upper limits in Fig.\ref{LrLx_sub} (left panel). The situation can be even complex thinking to the process of cluster formation and to the ensuing generation of cluster turbulence; cosmological simulations including a proper treatment of cosmic ray acceleration/cooling are necessary to shed light on these processes.

\section{Testing the re-acceleration scenario with low frequency observations}

The turbulent re-acceleration scenario explains the connection between RH and cluster mergers, the radio bi-modality of clusters and the fast evolution of clusters in the $P_{1.4}-L_X$ diagram.

\begin{figure}
\centering
\includegraphics[width=7cm, angle=0]{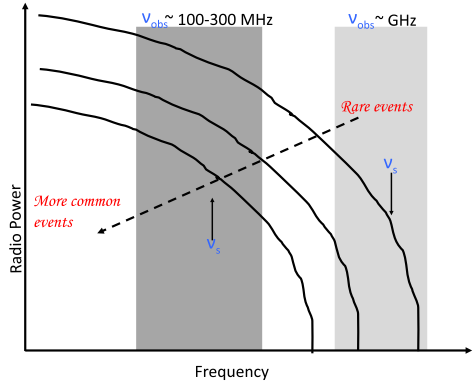}
\caption{\footnotesize
Schematic representation of the synchrotron spectra of RH with different values of $\nu_s$ (see text for details). Those with higher values 
of $\nu_s$ are visible up to GHz frequencies, while those with lower values would be observable only at lower frequencies.}
\label{spettri}
\end{figure}

It is important to point out that the peculiarity of this scenario is the fact that turbulent acceleration is a poorly efficient process; this has consequences on model expectations and allow for a prompt test of this scenario with future observations. Electrons can be accelerated only up to energies of $m_e c^2 \gamma_{max} \leq$ several GeV, entailing a high-frequency cut-off in the synchrotron spectra of RH, which marks the most important and unique expectation of this scenario (see Fig.~\ref{spettri}). The presence of this cut-off implies that the observed fraction of clusters with RH depends on the observing frequency. 
The steepening of the spectrum makes it difficult to detect RH at frequencies larger than the frequency $\nu_s$ where the steepening becomes severe. The frequency $\nu_s$ depends on the acceleration efficiency in the ICM, which in turns depends on the flux of MHD turbulence dissipated in relativistic electrons (\eg Cassano et al. 2006, Cassano et al. 2010b). 
Larger values of $\nu_s$ are expected in more massive clusters and in connection with major merger events. 
As a consequence, according to this model, present radio surveys at $\sim$ GHz frequencies can reveal only those RH generated during the most energetic merger events and characterized by relatively flat spectra ($\alpha \sim 1.1-1.5$) (see Fig.\ref{spettri}). 
These sources should represent the tip of the iceberg of the whole population of RH, since the bulk of cluster formation in the Universe occurs through less energetic mergers. Low frequency observations with new generation of radio telescopes 
(LOFAR, LWA) are thus expected to unveil the bulk of RH, including a population of RH which will be observable preferentially at low radio frequencies ($\nu \leq 200-300$ MHz). 
These RH, generated during less energetic but more common merger events, should have extremely steep radio spectra ($\alpha \gtsim 1.5-1.9$) when observed at higher frequencies; we defined these sources Ultra Steep Spectrum RH (USSRH). 
Possible prototypes of these RH are those found in Abell 521 ($\alpha\sim 2$, Brunetti et al. 2008) and 
in Abell 697 ($\alpha\sim 1.7$, Macario et al. 2010).

In the framework of the {\it turbulent re-acceleration} scenario, the existence of merging clusters without Mpc-scale radio 
emission (Fig.~\ref{LrLx_sub}, right panel; see also the case of Abell 2146 by Russell et al. (2011)), is not surprising for two main reasons. 
First, the expected lifetime of RH ($\sim$ Gyr) can be smaller than the typical time-scale of a merger, during which the cluster appears disturbed, implying that not all disturbed systems should host RH (\eg Brunetti et al. 2009). 

\noindent
Second, and most important, a fraction of disturbed systems may host RH with very steep radio spectra (USSRH), difficult to detect even at low frequencies if the observations are not sensitive enough. USSRH are mainly expected in disturbed clusters with masses $M_v \ltsim 10^{15}\,M_{\odot}$ in the local Universe, or in merging and massive clusters at higher redshift ($z\gtsim 0.4-0.5$; Cassano et al. 2010b). 
In line with this scenario, 3 out of the 4 outliers in Fig.\ref{LrLx_sub} have X-ray luminosity close to the lower boundary used to select the GMRT sample ($L_X = 5\times 10^{44}$ erg/sec), and the other is the cluster with the highest redshift in the GMRT sample ($z \simeq 0.42$). Interestingly, a deep GMRT follow-up at 325 MHz of one of the outliers in Fig.~\ref{LrLx_sub}, Abell 781,  has revealed the presence of a possible USSRH (Venturi et al. 2011), which need to be confirmed by future deeper low-frequency observations.
The recent case of Abell 2146 may also be in line with these hypotesis: it has a moderate X-ray luminosity ($L_{[0.1-2.4] {\mathrm keV}}\sim 6\times 10^{44}$ erg/s) and it is at $z\sim0.23$. Only 5 RH are presently known in clusters with the same (or lower) X-ray luminosity but are all at smaller redshifts, which could suggest that the increase of the inverse Compton losses of relativistic electrons with redshift ($\propto (1+z)^4$) contribute to disfavour the formation of RH emitting at GHz frequencies in these less massive clusters.

USSRH are expected to be less powerful than RH emitting at GHz frequencies (see Cassano 2010) and thus very sensitive 
low-frequency observations are necessary to catch them. 
The ideal instrument to search for USSRH is LOFAR (LOw Frequency ARray) that is already operating in the commissioning phase (\eg R\"ottgering et al. 2010). Monte Carlo procedures, that follow the process of cluster formation, the injection and dissipation of turbulence during cluster-cluster mergers and the ensuing acceleration of relativistic particles in the ICM, allowed to derive quantitatively the statistical properties of RH (\eg Cassano \& Brunetti 2005). The expectations based on these procedures were found consistent with present observational constraints (\eg Cassano et al. 2008) and were used to derive expectations for the planned LOFAR surveys. Accordingly to these predictions the {\it Tier 1} ``Large Area Survey'' at 120 MHz (see R\"ottgering et al. 2010), with an expected rms sensitivity of 0.1 mJy/beam, should greatly increase the number of known giant RH with the possibility to detect about 350 RH up to redshift $z \approx 0.6$ in the northern hemisphere with about half of these RH having very steep radio spectra ($\alpha\gtsim1.9$, Cassano et al. 2010b).

This implies that future LOFAR surveys will allow a powerful test of the merger-driven turbulence re-acceleration scenario for the origin of RH.

\section{Conclusions}
\label{sec:future}

We discussed the most recent statistical evidences demonstrating the connection between 
giant RH and cluster mergers and the ``transient'' nature of the RH phenomenon. 

\noindent
A step forward in this direction comes from the discovery that the radio bi-modality of clusters
has a correspondence in terms of dynamical state of the clusters: clusters with RH are found to 
be dynamically disturbed, while clusters without RH are more dynamically relaxed. 
These observational evidences suggest that RH form in galaxy clusters and live for a period of time during 
mergers, when the clusters appear dynamically disturbed, while at the later times, when the clusters
become dynamically relaxed, the synchrotron emission fades away. 
Two main ingredients may drive the evolution of the cluster synchrotron emission: the magnetic field
and the relativistic electrons. Faraday Rotation measurements and observations of depolarization of cluster galaxies
suggest that the magnetic field has a marginal role, and favour a scenario where the generation of RH is connected with
the acceleration of relativistic particles.

These facts are naturally understood in the framework of one 
of the proposed pictures put forward to explain the origin of giant RH, 
the merger-induced turbulence re-acceleration scenario (Brunetti et al. 2001, Petrosian 2001). 
The main expectation of this scenario, which is related to the poorly efficient nature of the
turbulent acceleration mechanism in the ICM, is the existence of a population of clusters
hosting RH with very steep radio spectra, USSRH.
These RH, that should glow up preferentially at low radio frequency, are hidden in massive clusters
undergoing minor merging and/or less massive clusters interested by major merger events. 

LOFAR is an ideal instrument to test these expectations, and it is expected to discover
$\sim 350$ RH in the {\it Tier 1} ``Large Area Survey'' at 120 MHz, half of which should be USSRH.

\section*{Acknowledgments} 
This work is partially supported by INAF under grants PRIN-INAF2008 and
PRIN-INAF2009 and by ASI-INAF under grant I/088/06/0.

\end{document}